\def\etal{{\it et al.}}
\def\ave#1{\langle#1\rangle}        

\def\etc{{\it etc.}}
\def\etal{{\it et al.}}
\def\ave#1{\langle#1\rangle}        

\def\etc{{\it etc.}}
\def\ie{{\it i.e.}}

\def\lsim{\mathrel {\vcenter {\baselineskip 0pt \kern 0pt
    \hbox{$<$} \kern 0pt \hbox{$\sim$} }}}
    \def\gsim{\mathrel {\vcenter {\baselineskip 0pt \kern 0pt
    \hbox{$>$} \kern 0pt \hbox{$\sim$} }}}

\documentstyle[epsfig,aps,prd,twocolumn]{revtex}

\def\etal{{\it et al.}}
\def\ave#1{\langle#1\rangle}        

\def\etc{{\it etc.}}
\def\ie{{\it i.e.}}

\def\lsim{\mathrel {\vcenter {\baselineskip 0pt \kern 0pt
    \hbox{$<$} \kern 0pt \hbox{$\sim$} }}}
    \def\gsim{\mathrel {\vcenter {\baselineskip 0pt \kern 0pt
    \hbox{$>$} \kern 0pt \hbox{$\sim$} }}}

\begin{document}

\title{Classical ``Dressing'' of a Free Electron in a Plane Electromagnetic
Wave}

\author{Kirk T.~McDonald}
\address{Joseph Henry Laboratories, Princeton University, Princeton, NJ 08544}

\author{Konstantine Shmakov}
\address{Department of Physics and Astronomy,
University of Tennessee, Knoxville, Tennessee 37996}

\date{Feb.~28, 1998} 

\maketitle

\begin{abstract}
The energy and momentum densities of the fields of a free electron in a plane
electromagnetic wave include interference terms that are the classical
version of the ``dressing'' of the electron the arises in a quantum analysis.
The transverse mechanical momentum of the oscillating electron is balanced
by the field momentum resulting from the interference between the driving wave 
and the static part of the
electron's field.  The interference between the wave and the oscillating
part of the electron's field leads to a longitudinal field momentum and
a negative field energy that compensate for the longitudinal momentum and
kinetic energy of the electron.  The interference terms are dominated by
the near zone, so that as the wave passes the electron by the latter
reverts to its energy and momentum prior to the arrival of the wave.

\end{abstract}

\section{Introduction}

The behavior of a free electron in a electromagnetic wave is one of the
most commonly discussed topics in classical electromagnetism.  Yet, several
basic issues remain to be clarified.  These relate to the question:
to what extent can net energy be transferred from an electromagnetic pulse
(such as that of a laser) in vacuum to a free electron?

These issues are made more complex by quantum considerations, including the role
of the ``quasimomentum'' of an electron that is ``dressed'' by an
electromagnetic wave \cite{Malkacomment}.

As a small step towards understanding of the larger issues, we consider a 
simpler question here.  The response of a free electron to a plane 
electromagnetic wave is oscillatory motion in the plane perpendicular to the 
direction of the wave, in the first approximation.  Thus, the electron has
momentum transverse to the direction of the wave.  However, the wave contains
momentum only in its direction, and the radiated wave contains no net
momentum (in the nonrelativistic limit).  How is momentum conserved in this
process?

The general sense of the answer has been given by Poynting \cite{Poynting},
who noted that an electromagnetic field can be said to contain a flux of
energy (energy per unit area per unit time) given by
\begin{equation}
{\bf S} = {c {\bf E} \times {\bf B} \over 4 \pi},
\label{eq1}
\end{equation}
in Gaussian units, where {\bf E} is the electric field, {\bf B} is the
magnetic field (taken to be in vacuum throughout this paper) and $c$ is the
speed of light.

Poincar\'e \cite{Poincare00} noted that this flow of energy can also be
associated with a momentum density given by
\begin{equation}
{\bf P}_{\rm field} = 
{{\bf S} \over c^2}  = {{\bf E} \times {\bf B} \over 4 \pi c},
\label{eq2}
\end{equation}
Hence, in the problem of a free electron in a plane electromagnetic wave
we are led to seek an electromagnetic field momentum that is equal and 
opposite to the mechanical momentum of the electron.

In this paper we demonstrate that indeed the mechanical momentum of the
oscillating electron is balanced by the field momentum in the interference
term between the incident wave and the static field of the electron.
We are left with some subtleties when we consider the interference between
the incident wave and the oscillating field of the electron.

\section{Generalities}

\subsection{Motion of an Electron in a Plane Wave}

We consider a plane electromagnetic wave that propagates in the $+z$ direction
of a rectangular coordinate system.  A fairly general form of this wave is
\begin{eqnarray}
{\bf E}_{\rm wave} & = & \hat{\bf x} E_x \cos(kz - \omega t)
                       - \hat{\bf y} E_y \sin(kz - \omega t), \nonumber \\
{\bf B}_{\rm wave} & = & \hat{\bf x} E_y \sin(kz - \omega t)
                       + \hat{\bf y} E_x \cos(kz - \omega t),
\label{eq3}
\end{eqnarray}
where $\omega = kc$ is the angular frequency of the wave, $k = 2\pi/\lambda$
is the wave number and $\hat{\bf x}$ is a unit vector in the $x$ direction,
\etc\ \ When either $E_x$ or $E_y$ is zero we have a linearly
polarized wave, while for $E_x = \pm E_y$ we have circular polarization.

A free electron of mass $m$ oscillates in this field such that its average
position is at the origin.  This simple statement hides the subtlety that
our frame of reference is not the lab frame of an electron that is initially
at rest but which is overtaken by a wave \cite{Kibble,Sarachik,tempaccel}.  
If the velocity of the oscillating electron is small, we can ignore the
${\bf v}/c \times {\bf B}$ force and take the motion to be entirely in the
plane $z = 0$.  Then, (also ignoring radiation damping) the  equation of motion 
of the electron is
\begin{equation}
m \ddot{\bf x} = e {\bf E}_{\rm wave}(0,t) =
e( \hat{\bf x} E_x \cos \omega t + \hat{\bf y} E_y \sin \omega t) .
\label{eq4}
\end{equation}
Using eq.~(\ref{eq3}) we find the position of the electron to be
\begin{equation}
{\bf x} = -{e \over m \omega^2} ( \hat{\bf x} E_x \cos\omega t
                       + \hat{\bf y} E_y \sin\omega t), 
\label{eq5}
\end{equation}
and the mechanical momentum of the electron is
\begin{equation}
{\bf p}_{\rm mech}= m \dot{\bf x} = {e \over  \omega} 
( \hat{\bf x} E_x \sin\omega t - \hat{\bf y} E_y \cos\omega t). 
\label{eq6}
\end{equation}
The root-mean-square (rms) velocity of the electron is
\begin{equation}
v_{\rm rms} = \sqrt{\ave{\dot x^2 + \dot y^2}} 
= {e \over m \omega} \sqrt{E_x^2 + E_y^2 \over 2}
= {e E_{\rm rms} \over m \omega c} c.
\label{eq7}
\end{equation}
The condition that the ${\bf v}/c \times {\bf B}$ force be small is then
\begin{equation}
\eta \equiv {e E_{\rm rms} \over m \omega c} \ll 1,
\label{eq8}
\end{equation}
where the dimensionless measure of field strength, $\eta$, is a Lorentz
invariant.  
Similarly, the rms departure of the electron from the origin is
\begin{equation}
x_{\rm rms} = {e E_{\rm rms} \over m \omega^2} = {\eta \lambda \over 2\pi}.
\label{eq9}
\end{equation}
Thus, condition (\ref{eq8}) also insures that the extent of the  motion of the 
electron is small compared to a wavelength, and so we may use the dipole
approximation when considering the fields of the oscillating electron.

In the weak-field approximation, we can now use (\ref{eq6}) for the velocity to
evaluate the second term of the Lorentz force:
\begin{equation}
e{{\bf v} \over c} \times {\bf B} = {e^2 (E_x^2 - E_y^2) \over 2 m \omega c}
\hat{\bf z} \sin 2\omega t.
\label{eq9a}
\end{equation}
This term vanishes for circular polarization, in which case the motion is
wholely in the transverse plane.  However, for linear polarization the 
${\bf v}/c \times {\bf B}$ force leads to
oscillations along the $z$ axis at frequency $2 \omega$, as first analyzed
in general by Landau \cite{Landau}.  For polarization along
the $\hat{\bf x}$ axis, the $x$-$z$ motion has the form of a 
``figure 8'', which for weak fields ($\eta \ll 1$) is described by
\begin{equation}
x = - {e E_x \over m \omega^2} \cos \omega t, \qquad
z = - {e^2 E_x^2 \over 8 m^2 \omega^3 c} \sin 2\omega t.
\label{eq40}
\end{equation}

If the electron had been at rest before the arrival of the plane wave, then
inside the wave it would move with an average drift velocity given by
\begin{equation}
v_z = {\eta^2/2 \over 1 + \eta^2/2} c,
\label{eq9b}
\end{equation}
along the direction of the wave vector, as first deduced by McMillan
\cite{McMillan}.  In the present paper we work in the frame in which
the electron has no average velocity along the $z$ axis.  Therefore, prior to
its encounter with the plane wave the electron had been moving in the negative
$z$ direction with speed given by (\ref{eq9b}).

\subsection{Field Momentum}

The fields associated with the electron can be regarded as the
superposition of those of an electron at rest at the origin plus those of
a dipole consisting of the actual oscillating electron and a positron at
rest at the origin.  Thus, we can write the electric field of the electron
as ${\bf E}_{\rm static} + {\bf E}_{\rm osc}$ and the magnetic field as
${\bf B}_{\rm osc}$, where the oscillating fields have the pure frequency
$\omega$ in the low-velocity limit.

The entire electromagnetic momentum density can then be written
\begin{equation}
{\bf P}_{\rm field} = 
{({\bf E}_{\rm wave} + {\bf E}_{\rm static} + {\bf E}_{\rm osc})  \times 
({\bf B}_{\rm wave} + {\bf B}_{\rm osc}) \over 4 \pi c}.
\label{eq10}
\end{equation}
However, in seeking the field momentum that opposes the mechanical momentum of
the electron, we should not include either of the self-momenta
${\bf E}_{\rm wave} \times {\bf B}_{\rm wave}$ or
$({\bf E}_{\rm static} + {\bf E}_{\rm osc}) \times {\bf B}_{\rm osc}$.
The former is independent of the electron, while the latter can be
considered as a part of the mechanical momentum of the electron according to
the concept of ``renormalization''.

We therefore restrict our attention to the interaction field momentum
\begin{equation}
{\bf P}_{\rm int} = {\bf P}_{\rm wave,static} + {\bf P}_{\rm wave,osc},
\label{eq11}
\end{equation}
where
\begin{equation}
{\bf P}_{\rm wave,static} = {{\bf E}_{\rm static} \times 
{\bf B}_{\rm wave} \over 4 \pi c}.
\label{eq12}
\end{equation}
and
\begin{equation}
{\bf P}_{\rm wave,osc} = { {\bf E}_{\rm wave} \times {\bf B}_{\rm osc} 
+ {\bf E}_{\rm osc} \times {\bf B}_{\rm wave} \over 4 \pi c}.
\label{eq13}
\end{equation}

We recall from eqs.~(\ref{eq6}) and (\ref{eq40})
 that the transversemechanical momentum of the oscillating
electron has pure frequency $\omega$.  Since the wave and the oscillating
part of the electron's field each have frequency $\omega$, the term
${\bf P}_{\rm wave,osc}$ contains harmonic functions of $\omega^2$, which
can be resolved into a static term plus ones in frequency $2\omega$.  Hence
we should not expect this term to cancel the mechanical momentum.  Rather,
we look to the term ${\bf P}_{\rm wave,static}$, since this has pure
frequency $\omega$.

\section{The Momentum ${\bf P}_{\rm wave,static}$}

The static field of the electron at the origin is, in rectangular
coordinates,
\begin{equation}
{\bf E}_{\rm static} = {e \over r^3} (x \hat{\bf x} + y \hat{\bf y} +
z \hat{\bf z}),
\label{eq14}
\end{equation}
where $r$ is the distance from the origin to the point of observation.
Combing this with eq.~(\ref{eq3}) we have
\begin{eqnarray}
{\bf P}_{\rm wave,static} & = & {e \over 4 \pi c r^3} 
\{ - \hat{\bf x} z E_x \cos(kz - \omega t) 
\nonumber \\
& &  + \hat{\bf y} z E_y \sin(kz - \omega t)
\label{eq15} \\
& & + \hat{\bf z} [ x E_x \cos(kz - \omega t) - y E_y \cos(kz - \omega t)] 
\}.
\nonumber
\end{eqnarray}
When we integrate this over all space to find the total field momentum,
the terms in $\hat{\bf z}$ vanish  as they are odd in either $x$ or $y$.
Likewise, after expanding the cosine and sine of $kz - \omega t$, the
terms proportional to $z \cos kz$ vanish on integration.  The
remaining terms are thus
\begin{eqnarray}
{\bf p}_{\rm wave,static} & = & \int_V {\bf P}_{\rm wave,static}\ 
\label{eq16} \\
& = & {e \over 4 \pi c} 
( - \hat{\bf x}  E_x \sin \omega t + \hat{\bf y}  E_y \cos\omega t) 
\int_V {z \sin kz \over r^3} 
\nonumber \\
& = & {e \over \omega} 
( - \hat{\bf x}  E_x \sin \omega t + \hat{\bf y}  E_y \cos\omega t) 
= - {\bf p}_{\rm mech}, \nonumber
\end{eqnarray}
after an elementary volume integration.  

It is noteworthy that the integration
is independent of any hypothesis as to the size of a classical electron.
Indeed, the integrand of (\ref{eq16}) can be expressed as
$\cos\theta \sin (kr\cos\theta)/r^2$ via the substitution $z = r\cos\theta$.
Hence, the integral over a spherical
shell is independent of $r$ for $kr \ll 1$, and significant
contributions to the integral occur for radii up to one wavelength of the
electromagnetic wave.  This contrasts with the self-momentum density of the
electron which is formally divergent; if the integration is cut off at a
minimum radius (the classical electron radius), the dominant contribution occurs
within twice that radius.

Thus, we have demonstrated the principal result of this paper.

\section{The Momentum ${\bf P}_{\rm wave,osc}$}

Several subtleties in the argument appear when we consider the other
interference term in the momentum density (\ref{eq11}).  For this we must
first display the electromagnetic fields of an oscillating electron.

\subsection{The Fields ${\bf E}_{\rm osc}$ and ${\bf B}_{\rm osc}$}

Since we restrict our attention to an electron that oscillates with amplitude
much less than a wavelength of the driving wave, and the electron attains
velocities that are much less than the speed of light, it is sufficient to
use the dipole approximation to the fields of the electron.  While these
fields are well known, they are typically presented in imaginary notation, of
which only the real part has physical significance.  This notation is very
useful for discussions in which only time-averaged behavior is of interest.
However, we wish to consider the details of momentum balance at an arbitrary
moment, and it is preferable to use purely real notation.

We begin by noting that the retarded vector potential of the oscillating
electron at a point {\bf r} at time $t$  can be written
\begin{eqnarray}
{\bf A}_{\rm osc}({\bf r},t) & = & 
{e \over c} {\dot{\bf x}(t' = t - r/c) \over r}
\label{eq17} \\
& = & -{e^2 \over m \omega c r} [\hat{\bf x} E_x \sin(kr - \omega t)
                           + \hat{\bf y} E_y \cos(kr - \omega t)],
\nonumber
\end{eqnarray}
using eq.~(\ref{eq5}) for the motion {\bf x} of the electron.  The oscillating
part of the scalar potential is obtained by integration of the Lorentz gauge
condition:
\begin{equation}
\nabla \cdot {\bf A}_{\rm osc} + {1 \over c} {\partial \phi_{\rm osc} \over
\partial t} = 0. 
\label{eq18}
\end{equation}
We find
\begin{eqnarray}
\phi_{\rm osc} & = & - {e^2 \over m \omega^2} \left\{
E_x \left[ {kx \over r^2} \sin(kr-\omega t) + {x \over r^3} \cos(kr-\omega t) 
\right] \right. \nonumber \\
& & \left.
+ E_y \left[ {ky \over r^2} \cos(kr-\omega t) - {y \over r^3} \sin(kr-\omega t) 
\right] \right\}.
\label{eq19}
\end{eqnarray}
The constant static potential is omitted in the above.  

The scalar potential
could also be deduced from the retarded potential of a moving charge.
Equation (\ref{eq19}) results on expanding the retarded distance to first
order in the field strength of the plane wave.

The electric and magnetic fields are, of course, found from the potentials via
\begin{equation}
{\bf B} = \nabla \times {\bf A} \qquad \mbox{and} \qquad
{\bf E} = - \nabla \phi - {1 \over c} {\partial {\bf A} \over \partial t}.
\label{eq20}
\end{equation}
The lengthy expressions for the rectangular components of the fields are
\begin{eqnarray}
B_{{\rm osc},x} & = & - {e^2 E_y \over m \omega^2} \left[ 
{k^2 z \over r^2} \sin(kr-\omega t) + {kz \over r^3} \cos(kr-\omega t) \right],
\nonumber \\
B_{{\rm osc},y} & = & - {e^2 E_x \over m \omega^2} \left[ 
{k^2 z \over r^2} \cos(kr-\omega t) - {kz \over r^3} \sin(kr-\omega t) \right],
\nonumber \\
B_{{\rm osc},z} & = &  {e^2 E_x \over m \omega^2} \left[ 
{k^2 y \over r^2} \cos(kr-\omega t) - {ky \over r^3} \sin(kr-\omega t) \right]
\label{eq21} \\
& + & {e^2 E_y \over m \omega^2} \left[ 
{k^2 x \over r^2} \sin(kr-\omega t) + {kx \over r^3} \cos(kr-\omega t) \right],
\nonumber 
\end{eqnarray}
and
\begin{eqnarray}
E_{{\rm osc},x} & = & - {e^2 E_x \over m \omega^2} \biggl[ 
\left( {3 k x^2 \over r^4} - {k \over r^2} \right) \sin(kr-\omega t) 
\nonumber \\
& & \qquad +
\left( {k^2 \over r} - {k^2 x^2 \over r^3} + {3 x^2 \over r^5} - {1 \over r^3}
\right) \cos(kr-\omega t) \biggr]
\nonumber \\
&  & - {e^2 E_y \over m \omega^2} \biggl[ 
{3 k x y \over r^4} \cos(kr-\omega t)
\nonumber \\
& & \qquad + 
\left( {k^2 x y \over r^3} - {3 x y \over r^5} \right) \sin(kr-\omega t) 
\biggr],
\nonumber \\
E_{\rm osc,y} & = & - {e^2 E_x \over m \omega^2} \biggl[ 
{3 k x y \over r^4} \sin(kr-\omega t)
\nonumber \\
& & \qquad - 
\left( {k^2 x y \over r^3} - {3 x y \over r^5} \right) \cos(kr-\omega t) 
\biggr]
\nonumber \\
& & - {e^2 E_y \over m \omega^2} \biggl[ 
\left( {3 k y^2 \over r^4} - {k \over r^2} \right) \cos(kr-\omega t) 
\label{eq22} \\
& & \qquad -
 \left( {k^2 \over r} - {k^2 y^2 \over r^3} + {3 y^2 \over r^5} - {1 \over r^3}
\right) \sin(kr-\omega t)
\biggr], \nonumber \\
E_{\rm osc,z} & = & - {e^2 E_x \over m \omega^2} \biggl[ 
{3 k x z \over r^4} \sin(kr-\omega t)
\nonumber \\
& & \qquad  - 
\left( {k^2 x z \over r^3} - {3 x z \over r^5} \right) \cos(kr-\omega t) 
\biggr]
\nonumber \\
& & - {e^2 E_y \over m \omega^2} \biggl[ 
{3 k y z \over r^4} \cos(kr-\omega t)
\nonumber \\
& & \qquad + 
\left( {k^2 y z \over r^3} - {3 y z \over r^5} \right) \sin(kr-\omega t) 
\biggr].
\nonumber 
\end{eqnarray}
These expressions can also be deduced from the Li\'enard-Wiechert forms for
the fields of an accelerated charge, keeping terms only to first order in
the strength of the plane wave.

\subsection{Components of ${\bf P}_{\rm wave,osc}$}

Since the wave fields have no $z$ component, the $x$ component of
${\bf P}_{\rm wave,osc}$ is given by
\begin{equation}
P_{{\rm wave,osc},x} =
{E_{{\rm wave},y} B_{{\rm osc},z} - E_{{\rm osc},z} B_{{\rm wave},y} \over
4 \pi c}.
\label{eq23}
\end{equation}
From eqs.~(\ref{eq21}) and (\ref{eq22}) we see that both $B_{{\rm osc},z}$ and
$E_{{\rm osc},z}$ are odd in either $x$ or $y$.  Therefore, the volume
integral of $P_{{\rm wave,osc},x}$ vanishes, and we do not consider it
further.  Likewise, $P_{{\rm wave,osc},y}$  vanishes on integration.
This confirms the claim made at the end of sec.~II that the interference term
${\bf P}_{\rm wave,osc}$ is not relevant to the balance of transverse momentum
between the electron and the fields.

However, the $z$ component of ${\bf P}_{\rm wave,osc}$ does not vanish on
integration, and requires further discussion.  As the details include some
surprises (to the author) I present them at length.
\begin{eqnarray}
& & P_{{\rm wave,osc},z} =
\nonumber \\
& & {E_{{\rm w},x} B_{{\rm o},y} - E_{{\rm w},y} B_{{\rm o},x} 
+ E_{{\rm o},x} B_{{\rm w},y} - E_{{\rm o},y} B_{{\rm w},x}  \over
4 \pi c} =
\nonumber \\
& - & {e^2 E_x^2 \cos(kz-wt) \over 4 \pi m \omega^2 c} \biggl[
{k^2 z \over r^2} \cos(kr - \omega t)
- {k z \over r^3} \sin(kr - \omega t) \biggr]
\nonumber \\
& - & {e^2 E_y^2 \sin(kz-wt) \over 4 \pi m \omega^2 c} \biggl[
{k^2 z \over r^2} \sin(kr - \omega t)
+ {k z \over r^3} \cos(kr - \omega t) \biggr]
\nonumber \\
& - & {e^2 E_x^2 \cos(kz-wt) \over 4 \pi m \omega^2 c} \biggl[
\left( {3 k x^2 \over r^4} - {k \over r^2} \right) \sin(kr-\omega t) 
\nonumber \\
& & \qquad +
\left( {k^2 \over r} - {k^2 x^2 \over r^3} + {3 x^2 \over r^5} - {1 \over r^3}
\right) \cos(kr-\omega t) \biggr]
\nonumber \\
& - & {e^2 E_x E_y \cos(kz-wt) \over 4 \pi m \omega^2 c} \biggl[
{3 k x y \over r^4} \cos(kr-\omega t)
\nonumber \\
& & \qquad + 
\left( {k^2 x y \over r^3} - {3 x y \over r^5} \right) \sin(kr-\omega t) 
\biggr]
\label{eq24} \\
& - & {e^2 E_x E_y \sin(kz-wt) \over 4 \pi m \omega^2 c} \biggl[
-{3 k x y \over r^4} \sin(kr-\omega t)
\nonumber \\
& & \qquad  + 
\left( {k^2 x y \over r^3} - {3 x y \over r^5} \right) \cos(kr-\omega t) 
\biggr]
\nonumber \\
& + & {e^2 E_y^2 \sin(kz-wt) \over 4 \pi m \omega^2 c} \biggl[
\left( {3 k y^2 \over r^4} - {k \over r^2} \right) \cos(kr-\omega t) 
\nonumber \\
& & \qquad -
 \left( {k^2 \over r} - {k^2 y^2 \over r^3} + {3 y^2 \over r^5} - {1 \over r^3}
\right) \sin(kr-\omega t)
\biggr]. \nonumber
\end{eqnarray}

The terms of $P_{{\rm wave,osc},z}$ that are proportional to $E_y E_y$ are odd
on both $x$ and $y$, and so will vanish on integration.

We now consider the implications of eq.~(\ref{eq24}) separately for waves of
circular and linear polarization.

\subsection{Circular Polarization}

For a circularly polarized wave, we have $E_x^2 = E_y^2$.  Consequently
the dimensionless measure of field strength is $\eta = eE_x/m \omega c =
eE_y/m\omega c$, according to (\ref{eq8}).  The prefactors $e^2 E_x^2 /
4 \pi m \omega^2 c$ and $e^2 E_y^2 / 4 \pi m \omega^2 c$ can therefore
both be written $\eta^2 m c / 4 \pi$, and have dimensions of momentum.

The terms of eq.~(\ref{eq24}) in $E_x^2$ and $E_y^2$ can be combined in
pairs via the identities
\begin{eqnarray}
& & \cos(kz - \omega t) \cos(kr - \omega t) + 
\sin(kz - \omega t) \sin(kr - \omega t)
\nonumber \\
& & = \cos kz \cos kr + \sin kz \sin kr, 
\label{eq25}
\end{eqnarray}
and
\begin{eqnarray}
& & \sin(kz - \omega t) \cos(kr - \omega t) -
\cos(kz - \omega t) \sin(kr - \omega t)
\nonumber \\
& & = \sin kz \cos kr - \cos kz \sin kr.
\label{eq26}
\end{eqnarray}
A detail: the second term of eq.~(\ref{eq24}) in $E_x^2$ contains factors of
$x^2$, while second term of in $E_y^2$ contains factors of
$y^2$.  But during integration, we can replace $y^2$ by $x^2$, after which
the terms can be combined via (\ref{eq25}-\ref{eq26}).

We see already that the volume integral of $P_{{\rm wave,osc},z}$ will 
contain no time dependence!

On integration, terms such as $f(x,r)\sin kz$ and $g(x,r)z \cos kz$ that 
are odd in $z$ will vanish.  The integrated field momentum is thus,
\begin{equation}
p_{{\rm wave,osc},z} = \int_V P_{{\rm wave,osc},z} = 
- {\eta^2 m c \over 4 \pi} I_1 = - {4 \over 3} \eta^2 mc,
\label{eq27}
\end{equation}
where $I_1$ is the volume integral whose integrand is
\begin{eqnarray}
& & {k^2 z \over r^2} \sin kz\ \sin kr + {kz \over r^3} \sin kz\ \cos kr
\nonumber \\
& & +
\left( {3 k x^2 \over r^4} - {k \over r^2} \right) \cos kz\ \sin kr
\nonumber \\
& & +
 \left( {k^2 \over r} - {k^2 y^2 \over r^3} + {3 y^2 \over r^5} - {1 \over r^3}
\right) \cos kz\ \cos kr.
\label{eq28}
\end{eqnarray}

We return to the significance of eq.~(\ref{eq27}) after describing the
evaluation of integral $I_1$.

As seen from eq.~(\ref{eq27}), the integral $I_1$ must be dimensionless,
although it is apparently a function of the wave number $k$.   
However, the form of (\ref{eq28}) indicates that $I_1$ is actually independent 
of the length scale, so we
can set $k =1$ during integration.

To perform the integration we consider a volume element
$r^2 dr\ d\cos\theta\ d\phi$ in a spherical coordinate system with angle
$\theta$ defined relative to the $z$ axis.  It is more convenient to
keep $z = r \cos\theta $ as a variable of integration, using $dz = r 
d\cos\theta$.  Then the volume integration has the form
\begin{equation}
\int_V = \int_0^\infty rdr \int_{-r}^r dz \int_0^{2\pi} d\phi.
\label{eq29}
\end{equation}

Most terms of (\ref{eq28}) are independent of $\phi$, so their $\phi$ integral
is just $2\pi$.  For the terms in $x^2$, we have
\begin{equation}
\int_0^{2\pi} x^2\ d\phi = \int r^2 \sin^2\theta \cos^2\phi\ d\phi
 = \pi (r^2 - z^2).
\label{eq30}
\end{equation}

While each of the four main terms of (\ref{eq28}) diverges on integration, it
turns out that the two terms in $\cos z$ taken together are finite (and likewise
for the two terms in $\sin z$).  We find that
\begin{equation}
I_1 = I_A + I_B = {16 \pi \over 3},
\label{eq31}
\end{equation}
where 
\begin{eqnarray}
I_A & = & 2 \pi \int_0^\infty dr\ {\sin r \over r} \int_{-r}^r dz\ z \sin z 
\nonumber \\
& & + 2 \pi \int dr\ {\cos r \over r^2} \int dz\ z \sin z 
\nonumber \\
& = & 4 \pi,
\label{eq32}
\end{eqnarray}
and
\begin{eqnarray}
I_B & = &  \pi \int dr\ {\sin r \over  r} \int dz\ \cos z 
\nonumber \\
& & - 3 \pi \int dr\ {\sin r \over r^3} \int dz\ z^2 \cos z 
\nonumber \\
& & +\pi \int dr\ \cos r \left( 1 + {1 \over  r^2} \right) \int dz\ \cos z 
\nonumber \\
& & + \pi \int dr\ {\cos r \over r^2} \left( 1 - {3 \over  r^2} \right) 
 \int dz\ z^2 \cos z 
\nonumber \\
& = & {4\pi \over 3}.
\label{eq33}
\end{eqnarray}

From detailed evaluation of the radial integral, we find that the integrand
approaches a constant value as $r$ goes to zero, and that the contribution
to the integral at large $r$ diminishes as $1/r$.  That is, the principal
contribution is from the region $kr \approx 1$.

We are left with the result (\ref{eq27}) that the integral of the
interference term in the field momentum density has a constant longitudinal
term for an electron oscillating in a circularly polarized wave.

Recall that we have performed the analysis in a frame in which the electron
has no longitudinal momentum.  However, as remarked in sec.~IIA, prior to
its encounter with the wave, the electron had velocity $v_z = - \eta^2 c/2$
(assuming $\eta^2 \ll 1$), and therefore had initial mechanical momentum
$p_{{\rm mech},z} = - \eta^2 mc/2$.  So, we would expect that
this initial mechanical momentum had been converted to field momentum, if
momentum is to be conserved.

The result (\ref{eq27}) can be described as a kind of ``hidden momentum''
\cite{Hnizdo},
whose appearance can be surprising if one ignores the physical processes
needed to arrive at the nominal conditions of the problem.

We continue to be puzzled as to why the result (\ref{eq27}) is 8/3 times larger
than that required to satisfy momentum conservation.

\subsection{Linear Polarization}

Consider now the case of a linearly polarized wave with electric field along
the $x$ axis.  Then $E_{\rm rms} = E_x/\sqrt{2}$, and the prefactors in 
(\ref{eq24}) can be written as $\eta^2 mc/2\pi$.

The remaining terms in the momentum density $P_{{\rm wave,osc},z}$ have
time dependences that can be expressed as sums of pure frequencies via
the identities
\begin{eqnarray}
& & 2 \cos(kz - \omega t) \cos(kr - \omega t)
\nonumber \\
& = & \cos kz \cos kr + \sin kz \sin kr 
\nonumber \\
& & + (\cos kz \cos kr - \sin kz \sin kr) \cos 2\omega t 
\label{eq34} \\
& & + (\cos kz \sin kr + \sin kz \sin kr) \sin 2\omega t,
\nonumber
\end{eqnarray}
and
\begin{eqnarray}
& & 2 \cos(kz - \omega t) \sin(kr - \omega t)
\nonumber \\
& = & \cos kz \sin kr - \sin kz \cos kr 
\nonumber \\
& & + (\cos kz \sin kr + \sin kz \cos kr) \cos 2\omega t 
\label{eq35} \\
& & + (\sin kz \sin kr - \cos kz \cos kr) \sin 2\omega t,
\nonumber
\end{eqnarray}

Inserting these into eq.~(\ref{eq24}) and keeping only those terms that are
even in $z$, we find the integrated field momentum to be
\begin{eqnarray}
p_{{\rm wave,osc},z} & = & \int_V P_{{\rm wave,osc},z} 
\nonumber \\
& = & - {\eta^2 m c \over 4 \pi} (I_1 + I_2 \cos 2\omega t 
+ I_3 \sin 2\omega t),
\label{eq36}
\end{eqnarray}
where integral $I_1 = 16\pi/3$ has been discussed in (\ref{eq28}-\ref{eq33}),
\begin{equation}
I_2 = - I_A + I_B = -{8\pi \over 3},
\label{eq37}
\end{equation}
and integral $I_3$ has the integrand,

\begin{eqnarray}
& & {k^2 z \over r^2} \sin kz\ \sin kr  - {kz \over r^3} \sin kz\ \cos kr
\nonumber \\
& & -
\left( {3 k x^2 \over r^4} - {k \over r^2} \right) \cos kz\ \sin kr
\nonumber \\
& & +
 \left( {k^2 \over r} - {k^2 y^2 \over r^3} + {3 y^2 \over r^5} - {1 \over r^3}
\right) \cos kz\ \cos kr.
\label{eq38}
\end{eqnarray}
On evaluation, $I_3 = 0$. 

Hence, the longitudinal component of the interference field momentum of a free
electron in a linearly polarized wave is
\begin{equation}
p_{{\rm wave,osc},z} =
- {4 \over 3} \eta^2 m c + {2 \over 3} \eta mc \cos 2\omega t.
\label{eq39}
\end{equation}
The constant term is the same as that found in eq.~(\ref{eq27}) for
circular polarization, and represents the initial mechanical momentum of
the electron that became stored in the electromagnetic field once the electron
became immersed in the wave.

As for the second term of (\ref{eq39}), recall from eq.~(\ref{eq40}) that for
linear polarization the electron oscillates along the $z$ axis at frequency
$2 \omega$.  Hence the $z$ component of the mechanical momentum of the 
electron is
\begin{equation}
p_{{\rm mech},z} = m \dot z = - {\eta^2 mc \over 2} \cos 2\omega t.
\label{eq41}
\end{equation}
The term in $p_{{\rm wave,osc},z}$ at frequency $2\omega$ is $-4/3$ of the
longitudinal component of the mechanical momentum associated with the
``figure 8'' motion of the electron.  Thus, we have not been completely
successful in accounting for momentum conservation when the small, oscillatory
longitudinal momentum is considered.  

The factors of 4/3 and 8/3 are presumably not the same as the famous factor of
$4/3$ that arise in analyses of the electromagnetic energy and 
momentum of the self fields of an electron \cite{Schwinger,Moylan}.
A further appearance of a factor of 8/3 in the present example occurs when
we consider the field energy of the interference terms.

\section{The Interference Field Energy}

It is also interesting to examine the electromagnetic field energy of an
electron in a plane wave.  As for the momentum density (\ref{eq10}), we can
write
\begin{equation}
U_{\rm total} = 
{({\bf E}_{\rm wave} + {\bf E}_{\rm static} + {\bf E}_{\rm osc})^2
+ ({\bf B}_{\rm wave} + {\bf B}_{\rm osc})^2 \over 8 \pi },
\label{eq42}
\end{equation}
for the field energy density.  Again, we no not consider the divergent
energies of the self fields, but only the interference terms,
\begin{equation}
U_{\rm int} = U_{\rm wave,static} + U_{\rm wave,osc},
\label{eq43}
\end{equation}
where
\begin{equation}
U_{\rm wave,static} = { {\bf E}_{\rm wave} \cdot {\bf E}_{\rm static} 
\over 4 \pi}.
\label{eq44}
\end{equation}
and
\begin{equation}
U_{\rm wave,osc} = { {\bf E}_{\rm wave} \cdot {\bf E}_{\rm osc} 
+ {\bf B}_{\rm wave} \cdot {\bf B}_{\rm osc} \over 4 \pi}.
\label{eq45}
\end{equation}

In general, the interference field energy density is oscillating.  Here, we 
look for terms that are nonzero after averaging over time.  We see at once
that
\begin{equation}
\ave{U_{\rm wave,static}} = 0,
\label{eq46}
\end{equation}
since all terms have time dependence of $\cos \omega t$ or $\sin \omega t$.
In contrast, $\ave{U_{\rm wave,osc}}$ will be nonzero as its terms are
products of sines and cosines:
\begin{eqnarray}
& & U_{\rm wave,osc} =
\nonumber \\
& - & {e^2 E_x^2 \cos(kz-wt) \over 4 \pi m \omega^2 } \biggl[
\left( {3 k x^2 \over r^4} - {k \over r^2} \right) \sin(kr-\omega t) 
\nonumber \\
& & \qquad +
\left( {k^2 \over r} - {k^2 x^2 \over r^3} + {3 x^2 \over r^5} - {1 \over r^3}
\right) \cos(kr-\omega t) \biggr]
\nonumber \\
& - & {e^2 E_x E_y \cos(kz-wt) \over 4 \pi m \omega^2 } \biggl[
{3 k x y \over r^4} \cos(kr-\omega t)
\nonumber \\
& & \qquad + 
\left( {k^2 x y \over r^3} - {3 x y \over r^5} \right) \sin(kr-\omega t) 
\biggr],
\nonumber \\
& + & {e^2 E_x E_y \sin(kz-wt) \over 4 \pi m \omega^2 } \biggl[
{3 k x y \over r^4} \sin(kr-\omega t)
\nonumber \\
& & \qquad -
\left( {k^2 x y \over r^3} - {3 x y \over r^5} \right) \cos(kr-\omega t) 
\biggr]
\label{eq47} \\
& + & {e^2 E_y^2 \sin(kz-wt) \over 4 \pi m \omega^2 } \biggl[
\left( {3 k y^2 \over r^4} - {k \over r^2} \right) \cos(kr-\omega t) 
\nonumber \\
& & \qquad -
 \left( {k^2 \over r} - {k^2 y^2 \over r^3} + {3 y^2 \over r^5} - {1 \over r^3}
\right) \sin(kr-\omega t)
\biggr]
\nonumber \\
& - & {e^2 E_y^2 \sin(kz-wt) \over 4 \pi m \omega^2 } \biggl[
{k^2 z \over r^2} \sin(kr - \omega t)
+ {k z \over r^3} \cos(kr - \omega t) \biggr]
\nonumber \\
& - & {e^2 E_x^2 \cos(kz-wt) \over 4 \pi m \omega^2 } \biggl[
{k^2 z \over r^2} \cos(kr - \omega t)
- {k z \over r^3} \sin(kr - \omega t) \biggr].
\nonumber
\end{eqnarray}

The terms in $E_x E_y$ will vanish on integration over volume.  The various
time averages are
\begin{eqnarray}
\ave{ 2 \cos(kz & - & \omega t) \cos(kr - \omega t)} 
\nonumber \\
& & = \cos kz \cos kr + \sin kz \sin kr, 
\nonumber \\
\ave{ 2 \sin(kz & - & \omega t) \cos(kr - \omega t)}
\nonumber \\
& & = \sin kz \cos kr - \cos kz \sin kr,
\nonumber \\
\ave{ 2 \cos(kz & - & \omega t) \sin(kr - \omega t)}
\nonumber \\
& & = \cos kz \sin kr - \sin kz \cos kr,
\nonumber \\
\ave{ 2 \sin(kz & - & \omega t) \sin(kr - \omega t)}
\nonumber \\
& & = \cos kz \cos kr + \sin kz \sin kr.
\label{eq48}
\end{eqnarray}

After performing the time average on eq.~(\ref{eq47}), we keep only terms that
are even in $z$.  These terms have the form (\ref{eq28}), and so we find that
\begin{equation}
u_{\rm int} = \int_V \ave{U_{\rm wave,osc}} 
= - {e^2 (E_x^2 + E_y^2) \over 8 \pi m \omega^2 } I_1
= - {4 \over 3} \eta^2 mc^2,
\label{eq49}
\end{equation}
for waves of either linear or circular polarization.
As with the case of the interference field momentum, this interference field
energy is distributed over a volume of order a cubic wavelength around the
electron.  Being an interference term, its sign can be negative.

We can interpret the quantity,
\begin{equation}
{u_{\rm int} \over c^2}= - {4 \over 3} \eta^2 m,
\label{eq50}
\end{equation}
as compensation for the relativistic mass increase of the oscillating electron,
which scales as $v_{\rm rms}^2/c^2$ and hence as $\eta^2$ (for small $\eta$, 
recall eq.~(\ref{eq7})).
Indeed, a general result for the motion of an electron in a plane wave of
arbitrary strength $\eta$ is
that its rms relativistic mass, often called its effective mass, is
\cite{Kibble,Landau}
\begin{equation}
m_{\rm eff} = m \sqrt{1 + \eta^2}.
\label{eq51}
\end{equation}
For small $\eta$, the increase in mass is
\begin{equation}
\Delta m \approx {1 \over 2} \eta^2 m.
\label{eq52}
\end{equation}

Thus, the decrease in field energy due to the
interference terms between the electromagnetic fields of the wave and electron
is $-8/3$ times the mass increase it should compensate.

\section{Discussion}

\subsection{Temporary Acceleration}

We remarked in sec.~IIA that the preceding analysis holds in the average
rest frame of the electron.  If instead the electron had been at rest prior
to the arrival of the plane wave, the velocity of the average rest frame
would be  $v_z = (\eta^2/2)/(1 + \eta^2/2)$.
For this, the amplitude of the plane wave is 
presumed to have a slow rise from zero to a long plateau at strength $\eta$,
followed by a slow decline back to zero.

Once the wave has passed by the electron, the interference field energy,
(\ref{eq49}), goes to zero since the integral is dominated by the 
contribution at distances of order a wavelength from the electron.
Hence, the electron's kinetic energy must return to zero (or to its initial
value if that was nonzero).  A plane wave, or more precisely, a long pulse
that is very nearly a plane wave, cannot transfer net energy to an
electron.  The acceleration of the electron from zero velocity to $v_z$ is
only temporary, {\it i.e.,} for the duration of the plane wave pulse.

This result was first deduced by di Francia \cite{Francia} and by Kibble
\cite{Kibble} by different arguments.

\subsection{The Radiation Reaction}

Our analysis of the energy balance of an electron in a plane wave is not
quite complete.  We have neglected the energy radiated by the electron.
Since the rate of radiation is constant (once the electron is inside the
plane wave), the total radiated energy grows linearly with time, and
eventually becomes large.  The interference energy, (\ref{eq49}), is
constant in time, and hence cannot account for the radiated energy.

More to follow.....


\section{Appendix: Li\'enard-Wiechert Fields}

As an alternative to the dipole approximation, we consider the use of
the Li\'enard-Wiechert potentials and fields of a moving electron.  We
have limited our analysis to the case of a weak plane wave ($\eta \ll $1),
for which the velocity of the electron is always small $(\beta = v/c \ll 1)$.
In this case we may approximate the time-dependent part of the fields of
the electron as proportional to the strength of the field of the plane
wave (proportional to $\eta$.  Then we find that 
the Li\'enard-Wiechert fields of the electron are the same as the fields
in the dipole approximation.

We can show this in two ways.  First, we verify that the Li\'enard-Wiechert
potentials reduce to eqs.~(\ref{eq17}) and (\ref{eq19}).  Second, we can verify
directly that the Li\'enard-Wiechert fields are the same as eqs.~(\ref{eq21})
and (\ref{eq22}).

The Li\'enard-Wiechert potentials are
\begin{equation}
\phi = \left[ {e \over R(1 - {\bf \beta} \cdot \hat{\bf n}} \right], \qquad
{\bf A} = \left[ {e {\bf \beta}
 \over R(1 - {\bf \beta} \cdot \hat{\bf n}} \right],
\label{eq100}
\end{equation}
where the electron is at postion {\bf x}, the observer is at {\bf r},
their separation is {\bf R} = {\bf r} $-$ {\bf x}, the unit vector
$\hat{\bf n}$ is ${\bf R}/R$, and the brackets, [\ ], indicate that quantities
within are to be evaluated at the retarded time, $t' = t - R/c$.

We work in the average rest frame of the electron.
In the weak-field approximation we ignore the longitudinal motion of the
electron, (\ref{eq40}), which is quadratic in the strength of the plane wave.
Then the velocity vector of the electron is
\begin{equation}
{\bf \beta}(t) = {e \over m\omega c} \left( \hat{\bf x} E_x \sin\omega t
- \hat{\bf y} E_y \cos\omega t \right),
\label{eq101}
\end{equation}
from eq.~(\ref{eq6}).  The retarded velocity is thus,
\begin{eqnarray}
[{\bf \beta}] & = & {\bf \beta}(t' = t - R/c)
\label{eq102} \\
& = & - {e \over m\omega c} \left( \hat{\bf x} E_x \sin(kR - \omega t)
+ \hat{\bf y} E_y \cos(kr - \omega t) \right).
\nonumber
\end{eqnarray}
Distance $R$ differs from $r$ because the electron's oscillatory motion takes 
it away from the origin.  However, the amplitude of the motion is proportional
to strength of the plane wave.  Hence, we may replace $R$ by $r$ in
eq.~(\ref{eq102}) with error only in the second order of field strength.

Since the vector potential includes a factor ${\bf \beta}$ in the numerator, 
we can replace
$R$ by $r$ and $1 - {\bf \beta} \cdot \hat{\bf n}$ by 1 in the first
order in the field strength of the plane wave. Thus,
\begin{equation}
{\bf A} = 
-{e^2 \over m \omega c r} \left( \hat{\bf x} E_x \sin(kr - \omega t)
                           + \hat{\bf y} E_y \cos(kr - \omega t) \right),
\label{eq103}
\end{equation}
in agreement with eq.~(\ref{eq20}).

In the scaler potential, we first bring ${\bf \beta}$ to the numerator:
\begin{equation}
\phi \approx {e[1 + {\bf \beta} \cdot \hat{\bf n}] \over [R]}. 
\label{eq104}
\end{equation}
Unit vector $[\hat{\bf n}]$ differs from unit vector $\hat{\bf r}$ due
to the oscillation of the electron, which is proportional to the field
strength of the plane wave.  
For the scalar potential, however, we must expand the factor $1/[R]$ to first
order in the field strength.  Now,
\begin{equation}
[R] = |{\bf r} - {\bf x}(t')| = \sqrt{r^2 - 2{\bf r} \cdot {\bf x}(t')
+ {\bf x}^2(t')},
\label{eq105}
\end{equation}
with
\begin{eqnarray}
{\bf x}(t') & \approx & -{e \over m\omega^2} \left( \hat{\bf x} E_x 
\cos\omega t' + E_y \cos\omega t' \right)
\label{eq106} \\
& \approx & -{e \over m\omega^2} \left( \hat{\bf x} E_x 
\cos\omega(kr - \omega t) - \hat{\bf y} E_y \cos(kr - \omega t) \right),
\nonumber
\end{eqnarray}
again approximating $R$ by $r$ in the arguments of the cosine and sine,
accurate to first order in the field strength.  Hence,
\begin{eqnarray}
{1 \over [R]} & \approx & {1 \over r} (1 + {\bf r} \cdot {\bf x(t')})
\label{eq107} \\
& \approx & {1 \over r} \left\{ 1 - e  {\left(
x E_x \cos(kr - \omega t) - y E_y \sin (kr - \omega t) \right) 
\over m \omega^2 r^2} \right\}.
\nonumber
\end{eqnarray}
Altogether,
\begin{eqnarray}
\phi & \approx & {e \over r} - {e^2 \over m \omega^2}
\left\{ E_x \left( {kx \over r^2} \sin(kr - \omega t) + {x \over r^3}
\cos(kr - \omega t) \right) \right.
\nonumber \\
& & \left. + E_y \left( {ky \over r^2} \cos(kr - \omega t) - {y \over r^3}
\sin(kr - \omega t) \right) \right\},
\label{eq108}
\end{eqnarray}
in agreement with eq.~(\ref{eq19}).

Similarly, we could proceed from the Li\'enard-Wiechert fields,
\begin{eqnarray}
{\bf E} & = & e \left[ { \hat{\bf n} - {\bf \beta} \over \gamma^2 
(1 - {\bf \beta} \cdot \hat{\bf n} )^3 R^2 } \right] 
 + {e \over c} \left[ { \hat{\bf n} \times \left\{ 
( \hat{\bf n} - {\bf \beta} ) \times \dot{\bf \beta} \right\} 
\over (1 - {\bf \beta} \cdot \hat{\bf n} )^3 R } \right],
\nonumber \\
{\bf B} & = & [ \hat{\bf n} \times {\bf E} ].
\label{eq109}
\end{eqnarray}
After some work, we find that these fields are the same as 
eqs.~(\ref{eq21}-\ref{eq22}), to first order in the strength of the plane
wave.


\vspace{-0.2in}     


\begin{thebibliography}{[99]}

\vspace{-0.7in}     

\bibitem{Malkacomment}
K.T.~McDonald,
``Comment on ``Experimental Observation of Electrons Accelerated in
Vacuum to Relativistic Energies by a High-Energy Laser'' by G.~Malka, 
E.~Lefebvre and J.L.~Miquel, Phys.\ Rev.\ Lett.\ {\bf 78}, 3314 (1997)'',
Phys.\ Rev.\ Lett.\ {\bf 80}, 1350 (1998).

\bibitem{Poynting}
J.H.~Poynting,
``On the Transfer of Energy in the Electromagnetic Field''
Phil.\ Trans.\ {\bf 175}, 343-361 (1884); also, pp.~174-193, {\sl Collected
Scientific Papers} (Cambridge U.\ Press, 1920). 

\bibitem{Poincare00}
H.~Poincar\'e,
``Th\'eorie de Lorentz et le Principe de la R\'eaction'',
Arch.\ Ne\'erl.\ {\bf 5}, 252-278 (1900).

\bibitem{Kibble}
T.W.B.~Kibble,
``Frequency Shift in High-Intensity Compton Scattering'', 
Phys.\ Rev.\ {\bf 138}, B740-753, (1965); 
``Radiative Corrections to Thomson Scattering from Laser Beams'', 
Phys.\ Lett.\ {\bf 20}, 627-628 (1966);
``Refraction of Electron Beams by Intense Electromagnetic Waves'',
Phys.\ Rev.\ Lett.\ {\bf 16}, 1054-1056 (1966);
`Mutual Refraction of Electrons and Photons'',
Phys.\ Rev.\ {\bf 150}, 1060-1069 (1966);
``Some Applications of Coherent States'', 
{\sl Carg\`ese Lectures in Physics}, Vol.~2, ed.\ by M.~L\'evy (Gordon and
Breach, New York, 1968), pp.~299-345.

\bibitem{Sarachik}
E.S.~Sarachik and G.T.Schappert,
``Classical Theory of the Scattering of Intense Laser Radiation by Free
Electrons'',
Phys.\ Rev.\ D {\bf 1}, 2738-2753 (1970).

\bibitem{tempaccel}
K.T.~McDonald and K.~Shmakov,
``Temporary Acceleration of Electrons While Inside an Intense Electromagnetic
Pulse'',
Phys.\ Rev.\ ST Accel.\ Beams {\bf 2}, 121301-121305 (1999).

\bibitem{Landau}
L.~Landau and E.M.~Lifshitz, 
{\sl The Classical Theory of Fields},
4th ed.\ (Pergamon Press, Oxford, 1975), prob.~2, \S~47 and prob.~2, \S~49; 
p.~112 of the 1941 Russian edition.

\bibitem{McMillan}
E.M.~McMillan,
``The Origin of Cosmic Rays'', 
Phys.\ Rev.\ {\bf 79}, 498-501 (1950).

\bibitem{Hnizdo}
V.~Hnizdo,
``Hidden momentum and the electromagnetic mass of a charge and current
carrying body'',
Am.\ J.\ Phys.\ {\bf 65}, 55-65 (1997), and references therein.

\bibitem{Schwinger}
J.~Schwinger,
``Electromagnetic Mass Revisited'',
Found.\ Phys.\ {\bf 13}, 373-383 (1983).

\bibitem{Moylan}
P.~Moylan,
``An Elementary Account of the Factor of 4/3 in the Electromagnetic Mass'',
Am.\ J.\ Phys.\ {\bf 63}, 818-820 (1995).  

\bibitem{Francia}
G.~Toraldo di Francia,
{\sl Interaction of Focused Laser Radiation with a Beam of Charged Particles},
Nuovo Cim.\ {\bf 37}, 1553 (1965).


\end{thebibliography}
\end{document}